\documentstyle[psfig,12pt,a4]{article}

\begin{document}%
\title{Berry's phase for compact Lie groups.}
\author{E. Strahov\footnote{strahov@physics.technion.ac.il}\\
{\sl Physics Dept., Technion-Israel Institute of Technology}\\
{\it Haifa, 32000 Israel}   }
\date{11.07.2000}
\maketitle
   \begin{abstract}  
The Lie group adiabatic evolution determined by a Lie algebra 
parameter dependent
Hamiltonian is considered. It is demonstrated that in the case 
when the parameter space of
the Hamiltonian is a homogeneous K\"ahler manifold its fundamental 
K\"ahler potentials completely determine Berry geometrical phase 
factor. Explicit expressions for Berry vector potentials
(Berry connections) and Berry curvatures are obtained 
using the complex parametrization of the Hamiltonian parameter 
space. 
A general approach is exemplified by the Lie algebra 
Hamiltonians corresponding to $SU(2)$ and $SU(3)$  evolution 
groups.\\
\vspace{1.5cm}
   \end{abstract}
\newpage
\section*{I. Introduction}
A number of branches of physics make use of the geometric properties of 
K\"ahler coset spaces (a definition may be found in Kobayashi and 
Nomizu, vol.2\cite{kobayashi}).
For instance, 
in quantum field theory the K\"ahler coset spaces give rise   to
a broad class of supersymmetric non-linear sigma models discussed in
Zumino\cite{zumino}, Alvarez-Gaum\' e and Freedman\cite{alvarez},
Bando, Kuramoto, Maskawa and Uehara\cite{bando} (among others). 
In quantization of dynamical systems with curved phase space with a 
non-trivial global geometry K\"ahler cosets serve  as a model of 
such curved phase space (e.g. Beresin\cite{beresin}, Bar-Moshe and 
Marinov\cite{marinov1},\cite{marinov2}). K\"ahler geometry is also 
used in relativity theory (for a review see Flaherty\cite{flaherty}). 

In this work the sphere of use of 
geometric properties of the K\"ahler coset spaces is extended to
usual non-relativistic quantum mechanics.
I show that knowledge of the fundamental K\"ahler potentials of these 
spaces enables to find a phase factor acquired by the quantum state 
under a compact group adiabatic evolution.  

Berry\cite{berry} was the first to discover the relation between the 
adiabatic phase acquired by the wave function under a slow variation 
of the Hamiltonian parameters and the geometry of the parameter
space. Specifically, it has been demonstrated that the adiabatic
 phase includes a part of a pure geometrical origin (the geometric
phase factor).
Simon\cite{simon} has shown that the geometrical meaning of the 
geometric phase is the holonomy in a Hermitian line bundle over
the parameter space of the Hamiltonian, and that the adiabatic 
theorem\cite{born} (see also Messiah\cite{messiah}) gives rise to a connection
 with such bundle. 
When the parameter dependence of the Hamiltonian is
determined by a closed curve $C$ on the parameter space, the Berry 
geometrical factor $\Omega$ is expressed by the 
integral (Simon\cite{simon}, Berry\cite{berry})
\begin{equation}\label{intro1}
\Omega(C)=\int\limits_{S}F.
\end{equation}
Here $S$ is any oriented surface in the parameter space with 
$\partial S=C$, and $F$ is a two-form given on this parameter
space. As a consequence of the Stock's theorem the two-form
$F$ may be expressed in terms of the Berry vector 
potential\cite{berry} (or Berry connection). 

However, explicit forms for the geometric phase factor $\Omega$
(Eq.\ref{intro1}) and for the Berry connection in terms of the 
(local)
coordinates of the parameter space have since been obtained only 
for a number of simple cases. 
The spin precession in a slowly 
time-dependent magnetic field when the parameter space is a 
two-dimensional sphere, and the Berry connection is expressed in 
terms of the spherical coordinates is the simplest example. 
  After a suitable reparametrization of the time 
variable the
 Hamiltonian $H$ for this case may be chosen as  
\begin{equation}\label{intro2}
H(s)= {\bf n}(s){\bf J}, \;\;\;
\left| {\bf n}(s)\right| =1,
\end{equation}
where ${\bf J}$ is the spin operator, and the vector 
${\bf n}=(\sin\theta\cos\phi, \sin\theta\sin\phi, \cos\theta)$.
(It has been shown by Jak\v si\'c and 
Segert\cite{jak} that any two-level system may be described by
the Hamiltonian (\ref{intro2}) 
(with ${\bf J}= \mbox{\boldmath $\sigma$}, \;\;
\mbox{\boldmath $\sigma$}=(\sigma_1, \sigma_2, \sigma_3)$
are Pauli matrices) after corresponding reparametrization of the 
time variable.)
The corresponing Berry connection $A_s$ induced by the adiabatic
evolution of the spin state is:
\begin{equation}\label{intro3}
A_s = \frac{l}{2}(1- \cos\theta)\dot\phi, \;\;\;
l=0, \pm 1, \pm 2 \dots .
\end{equation}
The Berry geometrical phase factor is defined by Eq.(\ref{intro1}), 
where the two-form $F$ is  
\begin{equation}\label{intro4}
F=\frac{l}{2} \sin\theta d\phi \wedge d\theta, \;\;\;
l=0, \pm 1, \pm 2 \dots .
\end{equation}
In this example the evolution operator acting on the spin state
belongs to an irreducible representation of the $SU(2)$ group,
the Hamiltonian $H(s)$ (Eq.\ref{intro2}) determines a smooth curve in 
the Lie algebra $su(2)$, and the parameter space is the homogeneous
space of the group $SU(2)$, $S^2=SU(2)/U(1)$.

It is a goal of the present paper to consider cases when the 
adiabatic evolution is determined by one-parameter Hamiltonians 
belonging to more complicated than $su(2)$ Lie algebras. 
I shall concentrate on one-parameter Hamiltonians which 
lead to compact group evolution operators, and 
determine closed
smooth curves in the semi-simple Lie algebras of arbitrary ranks.
The aim is to generalize equations Eq.(\ref{intro3}),
Eq.(\ref{intro4}), and to find  explicit expressions for Berry
connections and Berry curvatures in terms of the coordinates of
the corresponding parameter spaces.

This paper is organized as follows.
Section II starts with an introduction of  the relevant Lie algebra 
notations.
We first obtain that
the adiabatic phase factor is a scalar product of two vectors
in the root space of the Lie algebra under consideration.
We reveal that it is not always the case that 
the adiabatic phase factor  depends on one integer only
(as in the case of $su(2)$ Lie algebra, Eqs.(\ref{intro3}), 
(\ref{intro4})). Rather, it
is dependent on a set of integers with a number equal to the
rank of the Lie algebra (e.g. in the case of $su(2)$ the 
rank is equal to one, thus only one integer suffices). 
We note that these integers determine irreducible 
representation in which quantum states form a basis.

In Section III  a complex parametrization of the parameter space by a 
Mackey-type decomposition for any element of the evolution group ${\sf G}$
 is introduced. Applicability of this procedure 
is restricted to the cases when the parametric space of the 
Hamiltonian  is a homogeneous K\"ahler manifold
${\sf G/H}$. Since in the cases under consideration this 
restriction is always satisfied, it becomes possible to apply 
this method to find explicit expressions for the 
adiabatic phase factor and the Berry potential in terms of 
the coordinates of the parameter space (Section IV).
We discover that the Berry potentials may be 
expressed in terms of the fundamental K\"ahler potentials of 
the homogeneous K\"ahler manifold ${\sf G/H}$. 
Bando, Kuramoto, Maskawa, and Uehara's\cite{bando} method is
used to express the fundamental K\"ahler potentials in terms of 
the coordinates of the parameter space. Thus, 
the explicit expressions for Berry connections in terms of the 
complex parameters are found. This result will be formulated as a theorem
in Section IV. 
It will be demonstrated in Section 5 that the action of the group
${\sf G}$ on the K\"ahler manifold ${\sf G/H}$ induces the gauge
transformation of the Berry potentials.  Once explicit forms for the Berry 
connections are obtained, the Berry curvature and the Berry geometrical phase
factor are easily derived (Section V).

Specific cases of $SU(2)$ and $SU(3)$  evolution groups are 
considered in Section VI. Section VII concludes the paper.
\section*{II. Preliminaries and notations}
Assume that a matrix irreducible representation of a compact semi-simple
Lie group {\sf G} of order $n$ and rank $r$ is given. Let ${\cal G}$ be the Lie
algebra of {\sf G} in which ${\cal H} \in {\cal G}$ denotes its Cartan 
subalgebra. A canonical Cartan-Weyl basis 
 $\{h_j, e_{\alpha}, e_{-\alpha}\}$
in ${\cal G}$ is introduced ,
where $j=1, \dots, r \equiv \mbox{rank}\;\;{\cal G}$, and 
$\{\alpha\}\in {\bf \Delta}^+_{\cal G}$  are the positive roots of 
${\cal G}$. (The definitions and properties of semi-simple Lie algebras
 and Lie groups may be found in Gilmore\cite{gilmore}.)
The number of the positive roots is $n_+=\frac{1}{2}\left[n-r \right]$.
The canonical basis of the Lie algebra ${\cal G}$ may be chosen so that the 
commutation relations will be written in the following standard form:
\begin{eqnarray}\label{bg1}
\left[h_i h_j\right]=0 &,& [h_i, e_{\alpha}]=\alpha_i e_{\alpha}
\nonumber\\
\left[e_{\alpha}, e_{-\alpha}\right]=\sum\limits^{r}_{j=1} \alpha_i h_i &,& 
[e_{\alpha}, e_{\beta}]=\chi( \alpha, \beta)e_{\alpha +\beta}.
\end{eqnarray}
Here $\chi(\alpha, \beta)$ is a function on the root lattice which vanishes if
$\alpha +\beta \not\in \Delta^+_{\cal G}$. 
Choosing primitive roots $\mbox{\boldmath $\gamma$}_j, \; j=1, \dots, r$,
 the fundamental weights $\mbox{\boldmath $\omega$}_j, \; j=1, \dots, r$ are
determined from the equation
\begin{equation}\label{bg1a}
(\mbox{\boldmath $\omega$}_i \mbox{\boldmath $\gamma$}_i)= 
\frac{\delta_{ij}}{2}
(\mbox{\boldmath $\gamma$}_i \mbox{\boldmath $\gamma$}_i).
\end{equation}
For any unitary irreducible group
representation its dominant weight ${\bf l}$ is given by a sum of the 
fundamental weights with nonnegative integer coefficients:
\begin{equation}\label{bg2}
{\bf l}=\sum\limits^{r}_{j=1}l_j \mbox{\boldmath $\omega$}_j=
\sum\limits^r_{j=1}\tilde l_j {\bf x}_j,
\end{equation}
where $(\tilde l_1, \dots , \tilde l_r)$ are the coordinates of the
dominant weight vector  ${\bf l}$
in the Lie algebra root space in
which an orthogonal coordinate system is chosen. (Here and 
afterwards bold face is used to denote vectors in the Lie algebra root 
space.) The set 
$\{{\bf x}_j, j=1, \dots, r \}$ denotes the unit basis vector
of this coordinate system.

We are interested in the cyclic adiabatic evolution  of the dominant
weight vector
eigenket 
  $\psi_{\bf l}$ which is defined by the following equation:
\begin{equation}\label{bg2a}
h_j\psi_{\bf l}=\tilde l_j \psi_{\bf l},\;\;j=1,\dots ,r
\end{equation}
(For convenience we are dealing with
the dominant weight vector eigenket  $\psi_{\bf l}$, corresponding
to the dominant weight vector ${\bf l}$ here. The adiabatic
evolution of an arbitrary weight vector eigenket
may be considered in the same manner.)
This adiabatic evolution will be determined  by the Schr\"odinger equation
with a Lie algebra  Hamiltonian 
 $b(s)\in {\cal G}$ given in the irreducible representation 
$(l_1, \dots , l_r)$  of ${\cal G}$:
\begin{equation}\label{bg3}
i\dot\psi(s)=\tau b(s)\psi(s), \;\; \psi(0)=\psi_{\bf l}.
\end{equation}
(The physical time $t$ is replaced here by the scale time
 $s=t/\tau, \;s\in[0,1]$.
The adiabatic limit is
$\tau \rightarrow\infty$. The Hamiltonian 
$b(s)$ is assumed to  depend smoothly on $s\in [0,1]$.) 
The cyclic evolution means that $b(s)$ takes the same values at the ends of the
segment $[0, 1]$. In order for the initial state $\psi_{\bf l}$  defined by
Eq.(\ref{bg2a}) to be an eigenstate of the Hamiltonian $b(s)$, we demand  that
 $b(0)=b(1) \in {\cal H}$.

 The problem (\ref{bg3}) 
can be written in terms of the Cartan-Maurer one form:
\begin{equation}\label{bg5}
dgg^{-1}=-i\tau b(s)ds, \;\; g(s)\in {\sf G}, \;\; b(s)\in {\cal G}.
\end{equation}
Here $g(s)$ is the Lie group evolution operator in the irreducible 
representation
\linebreak $(l_1,\dots, l_r)$, $\psi(s)= g(s)\psi_{\bf l}$ and $g(0)=e$ is the unit
element of the compact evolution group ${\sf G}$. 
Geometrically, the given Lie algebra Hamiltonian $b(s)$
determines a closed smooth curve  in the Lie algebra ${\cal G}$ which 
begins and
finishes in the Cartan subalgebra ${\cal H}$ of ${\cal G}$. To solve
 Eq.(\ref{bg5}) means to 
find the corresponding curve on the group manifold ${\sf G}$. 

For any given $s$ the Hamiltonian $b(s) \in {\cal G}$ may be reduced  to the
Cartan subalgebra ${\cal H}$,
\begin{equation}\label{bg6}
b(s)=g_1(s)\beta(s)g^{-1}_1(s), \;\; \beta(s)\in {\cal H}.
\end{equation}
It is usefull to assume that
the Cartan subalgebra element $\beta$ does not
depend on the parameter $s\in[0,1]$, i.e.
the eigenvalues  of  the Hamiltonian $b(s)$
are constants on the segment under considerations.
For example, the $su(2)$ Lie algebra Hamiltonian given by 
Eq.(\ref{intro2}) has two constant eigenvalues $\pm 1$
if ${\bf J}= \mbox{\boldmath $\sigma$}$, ($\sigma_1, \sigma_2,
\sigma_3$) are Pauli matrices.
When $b(s)$ determines the closed curve that begins and finishes in ${\cal H}$,
$g_1(0)=g_1(1)=e$.
Assuming that the conditions of the adiabatic theorem  are satisfied, the 
probabilities of transitions induced by a
parameter dependence of $b(s)$ are supressed and the initial and final quantum
states are distinct in the phase factor only, the unknown group element $g(s)$
is decomposed as
\begin{equation}\label{bg8}
g(s)=g_1(s)h(s), \;\; h(s)\in {\sf H}.
\end{equation}
(${\sf H}\in {\sf G}$ denotes the Cartan subgroup of the Lie group {\sf G}). 
Inserting the decomposed  elements $b(s)$ (Eq.\ref{bg6}) and 
$g(s)$ (Eq.\ref{bg8}) into Eq.(\ref{bg5}) we obtain:
\begin{equation}\label{bg9}
dhh^{-1}=-i\tau \beta-g^{-1}_1dg_1. 
\end{equation}
The unknown Cartan subgroup element $h(s)\in {\sf H}$ depends on $r$ real 
parameters $\Theta^j(s)$,
\begin{equation}\label{bg10}
h(s)=\exp \left(-i \sum\limits_{j=1}^{r}\Theta^j(s)h_j\right).
\end{equation}
The parameters $\Theta^j(s)$ are obtained from Eq.(\ref{bg9}) using the 
orthogonality condition for the Cartan subalgebra basis elements 
${\rm Tr}(h_ih_j)=\delta_{ij}$. After a cyclic evolution $(s=1)$ we find
\begin{equation}\label{bg11}
\Theta^j(1)=\tau\beta^j-i\int\limits_0^1{\rm Tr}
\left(g^{-1}_1\frac{dg_1}{ds}h_j\right)ds.
\end{equation}
The second  term in  the above  expression  is  real
and  will determine the geometric part of  the adiabatic phase
acquired by  the dominant weight vector eigenket $\psi_{\bf l}$
after  the cyclic adiabatic evolution. 

Let us introduce the $r$-dimensional vectors in the root space 
of the Lie algebra ${\cal G}$:
\begin{equation}\label{bg11a}
\mbox{\boldmath $\beta$}=
\sum\limits_{j=1}^r\beta^j{\bf x}_j,\;\;
\mbox{\boldmath $\cal Q$}=
\sum\limits_{j=1}^r{\cal Q}^j{\bf x}_j.
\end{equation} 
The components ${\cal Q}^j$ of the vector
$\mbox{\boldmath $\cal Q$}$ are defined by the second 
term in Eq.(\ref{bg11}), i.e.
\begin{equation}\label{bg11b}
{\cal Q}^j =-i\int\limits^1_0{\rm Tr}\left(g^{-1}_1
\frac{dg_1}{ds}h_j \right) ds.
\end{equation}
After a cyclic adiabatic evolution the dominant weight vector  
eigenket $\psi_{\bf l}$ is multiplied
by the element of  the Cartan subgroup 
$h(1)$ given by Eqs.(\ref{bg10}),(\ref{bg11}),
i.e.
\begin{equation}\label{bg13}
\psi_{\bf l}\rightarrow h(1)\psi_{\bf l}=
\exp\left(-i \sum\limits_{j=1}^{r}\Theta^j(1)h_j\right)\psi_{\bf l}.
\end{equation}  
Using Eq.(\ref{bg2a}) we finally obtain
that the adiabatic phase factor $\Theta$ acquired by the quantum state
$\psi_{\bf l}$ is given by the scalar product
of the dominant weight vector ${\bf l}$ 
(Eq.\ref{bg2}) on the sum of the vectors 
$\mbox{\boldmath $\cal Q$}$ and $\mbox{\boldmath $\beta$}$:
\begin{equation}\label{bg14}
\psi_{\bf l}\rightarrow \exp\left(-i\Theta\right)\psi_{\bf l}, \;\;
 \Theta={\bf l}\cdot \mbox{\boldmath $\beta$}+{\bf l}\cdot 
\mbox{\boldmath $\cal Q$}.
\end{equation}
While the first term in the above expression for the adiabatic
phase factor $\Theta$ is associated with  the dynamical phase,
the second term ${\bf l}\cdot \mbox{\boldmath $\cal Q$}$ is the geometrical
 phase factor $\Omega$, defined as  
\begin{equation}\label{bg14a}
\Omega \equiv {\bf l}\cdot \mbox{\boldmath $\cal Q$}
\end{equation}
As we can see from Eq.(\ref{bg14}), the Berry geometrical phase factor depends
on integers $l_1, \dots, \l_r$ which determine the dominant weight
vector ${\bf l}$ (Eq.\ref{bg2}), and characterize the irreducible
representation of the evolution group under consideration.
\section*{III. Complex parametrization of the Hamiltonian parameter
space.}\label{section2}
The given Lie algebra Hamiltonian $b(s)\in {\cal G}$ varies adiabatically 
through a circuit $C$ in the parameter space which is the homogeneous 
group manifold ${\sf G/H}$. Indeed, $b(s)$ depends on $s$ only through 
the group element $g_1(s)$ (Eq.\ref{bg6}). The Hamiltonian 
$b(s)$ decomposition Eq.(\ref{bg6}) is invariant under $g_1(s)\rightarrow 
g_1(s)h_1, \;\;\forall h_1 \in {\sf H}$, so $g_1(s)$ must be chosen as a 
representative of the corresponding equivalence class. 
Thus, there is a gauge freedom in the diagonalization process
(Eq.\ref{bg6}), and the Cartan subgroup ${\sf H}\in {\sf G}$ is
the group of the gauge transformations.
Geometrically, ${\sf G}$ is described as a principal fiber bundle with 
the Cartan subgroup ${\sf H}$ as the standard fiber and ${\sf G/H}$ is 
the base coset space. 
According to Borel's theorem\cite{borel}, the necessary and sufficient
condition for the coset space ${\sf G/H}$ (where ${\sf G}$ is a compact
semi-simple group, and ${\sf H}$ is a closed subgroup of ${\sf G}$) to be
a homogeneous K\"ahler manifold is that ${\sf H}$ be the centrelizer
of a torus in ${\sf G}$. A torus means a direct product of any $U(1)$
subgroup of ${\sf H}$ and the centralizer means a subgroup which
consists of all  ${\sf G}$ elements commutative with that torus elements.
As it may be seen from Section 2, in the case under consideration 
conditions of the Borel theorem are satisfied (${\sf H}$  is a Cartan
subgroup commuting with a torus), so the Hamiltonian parameter is a 
homogeneous K\"ahler manifold. Then the complex 
parametrization on ${\sf G/H}$ may be introduced, and the explicit 
expression for the geometrical factor $\Omega$ may be found in terms of 
the (complex) coordinates of the parameter space ${\sf G/H}$. 

The desired complex parametrization on the homogeneous
group manifold ${\sf G/H}$
is introduced by the complex parametrization of 
the group element $g_1$, which determines decomposition of the 
Hamiltonian $b(s)$ (Eq.\ref{bg6}).
Namely, given the canonical basis, the Lie algebra ${\cal G}$ is split 
into three subalgebras, ${\cal G}={\cal H} \oplus {\cal B}_+ \oplus 
{\cal B}_-$, (${\cal B}_+, {\cal B}_-$ are called Borel subalgebras), 
corresponding to three subsets of the basis elements 
$\{e_{-\alpha}\}, \{h_j\}, \{e_{\alpha}\}$.
Respectively, the Lie algebra ${\cal B}_+ ({\cal B}_-)$ generates 
a nilpotent Borel subgroup 
${\sf B}_+ ({\sf B}_-) \subset {\sf G}^c$ (${\sf G}^c$ means
the complexification of the group ${\sf G}$). 
The element $g_1$ has a unique (left) Mackey 
decomposition 
\begin{equation}\label{bg15}
g_1 =ug_-, \;\; u\in {\sf B}_+, \;\; g_- \in {\sf G/B}_+.
\end{equation}
(Note that in order to get $u$ for any given $g_1$ one has to impose the 
condition that $g_-=u^{-1}g_1$ has no part in ${\sf B}_+$. That would 
determine $u$ completely). 
The complex parameters which can be introduced in ${\sf G/H}$ correspond 
to the positive roots of ${\cal G}$,
\begin{equation}\label{bg16}
u(z)=\exp \left(\sum\limits_{\alpha \in\Delta^+_{\cal G}} 
z^{\alpha}e_{\alpha}\right), \;\; z^{\alpha}\in {\cal C}.
\end{equation}
$u(z)$ is an element of a nilpotent group and its matrix representions
are polynominals of $z^{\alpha}$. The local form (\ref{bg16}) for $u(z)$ 
is valid in a neighborhood of the point $z^{\alpha}=0$, i.e. the origin 
of the coordinate system in ${\sf G/H}$. The origin is related to the 
choice of coordinates. A transition to other domains of ${\sf G/H}$ 
covering the K\"ahler manifold ${\sf G/H}$ may be performed by the group 
transformation.

Given $u(z)$, $g_-(z, \bar z)$ will acquire the following form:
\begin{equation}\label{bg17}
g_-(z, \bar z) =v^+(z, \bar z) k(z, \bar z), \;\; 
v(z, \bar z)\in {\sf B}_+,\;\; 
k(z, \bar z) \in {\sf H}.
\end{equation}
The elements $v^+(z, \bar z), k(z,\bar z)$ are expressed as exponentials 
of the corresponding Lie algebra elements:
\begin{equation}\label{bg18}
v^+(z, \bar z)=
\exp \Big( \sum\limits_{\alpha\in\Delta^+_{\cal G}} 
y^{\alpha}(z, \bar z)e_{-\alpha}\Big), \;\;
k(z,\bar z)= 
\exp\Big(\sum\limits^r_{i=1}\kappa^i(z, \bar z)h_i\Big).
\end{equation}
For a particular $u(z)\in {\sf B}_+$ the functions $y^{\alpha}(z, \bar 
z)$ and $\kappa^i(z, \bar z)$ may be determined when the group element 
$g_1$ is unitary:
\begin{equation}\label{bg19}
g^+_1 = g_1^{-1} \rightarrow v^+ kk^+ v =(u^+ u)^{-1}.
\end{equation}
$v$ is obtained from the (right) Mackey decomposition of $(u^+u)^{-1}$, 
and the explicit forms for the functions $y_{\alpha}(z,\bar z)$ may be 
found.  As soon as $v$ is given, one turns to calculation of $k$ from
the equation:
\begin{equation}\label{bg20}
kk^+ =(vu^+ uv^+)^{-1}\in {\sf H}
\end{equation}
The functions $\kappa^i(z,\bar z)$ are especially important as we shall 
see below. 
It will be shown in Section 4 that the functions $\kappa^i(z, \bar z)$
completely determine the Berry potentials when ${\sf G}$ is a compact
evolution group, and the Hamiltonian parameter space is ${\sf G/H}$.
These functions are lineary related with the fundamental 
K\"ahler potentials $K^i(z, \bar z)$ of the K\"ahler manifold ${\sf G/H}$ 
under considerations:
\begin{equation}\label{bg21}
K^i(z, \bar z)= -2\sum\limits^r_{j=1}\kappa^j(z,\bar z) {\rm Tr}(h_j \eta_i).
\end{equation}
The formula (\ref{bg21}) was obtained by Itoh, Kugo and 
Kunimoto\cite{itoh}. Here $\eta_i$ are the projection matrices introduced by 
Bando, Kuratomo, Maskawa and Uehara\cite{bando}. The projection matrices 
exist in any matrix representation of ${\sf G}$ and correspond to 
elements of the Cartan subalgebra $h_j \in {\cal H}$. The basic properties
of the projection matrices are \cite{bando}:
\begin{eqnarray}\label{bg22}
\eta_j=\eta^+_j \quad \eta^2_j =\eta_j \quad \eta_j \hat h_k=\hat 
h_k\eta_j \quad
\forall j, k=1, \dots , r \nonumber\\
\eta_j\hat e_{-\alpha}\eta_j=\hat e_{-\alpha} \eta_j \qquad
\eta_j \hat e_{\alpha}\eta_j = \eta_j \hat e_{\alpha} \qquad \qquad
\end{eqnarray}
(The hat stands for the matrix representation.) All $\eta_j$ are 
commuting with each other. For any
representation of ${\sf G}$, where $\hat h_j$ are diagonal, all $\eta_j$ 
are also diagonal. 
The explicit forms of $\eta_j$ satisfying Eq.(\ref{bg22}) may be found 
(Bando, Kuratomo, Maskawa and Uehara\cite{bando}).
For the irreducible representation under consideration the functions
$\kappa^j(z, \bar z), \; j=1, \dots, r$ may be expressed lineary in
terms of the fundamental K\"ahler potentials $K^j(z, \bar z), \;\;
j=1, \dots, r$ (Eq.\ref{bg21}). In its turn a suitable method of construction
of the fundamental K\"ahler potentials is given by Bando, Kuratomo, 
Maskawa and Uehara\cite{bando}( see also 
Itoh, Kugo and Kunitomo\cite{itoh}). A number of particular examples is 
considered by Marinov and Bar-Moshe\cite{marinov1}, \cite{marinov2} 
in relation to the geometric 
quantization on homogeneous compact K\"ahler manifolds. 

A technique for
constructing the fundamental K\"ahler potentials may be described as 
follows. Once the projection matrices $\eta_j$ are obtained 
from Eqs.(\ref{bg22}),
the projected determinant is defined for any matrix $M$ as
\begin{equation}\label{bg23a}
{\rm det}_{\eta_j} M \equiv {\rm det}(\eta_j M\eta_j + I-\eta_j).
\end{equation}
For any projection matrix $\eta_j$, a fundamental K\"ahler potential 
$K^j(z, \bar z)$ is constructed from the fundamental represenation
for the element $u(z)$ (Eq.\ref{bg16}) of the nilpotent Borel subgroup
${\sf B_+}$,
\begin{equation}\label{bg23b}
K^j(z, \bar z) \equiv \ln{\rm det}_{\eta_j}\Big(u(z)^+ u(z)\Big).
\end{equation}
Note that the fundamental K\"ahler potential $K^j(z, \bar z)$ is not a
global function on ${\sf G/H}$, except for cases where ${\sf G/H}$ has a
trivial topology. However, the manifold ${\sf G/H}$ may be covered with 
complex coordinate neighborhoods. A transition from one neighborhood to 
another may be given by the group transformation. If the group ${\sf G}$
acts holomorphically on ${\sf G/H}$,
$z \rightarrow gz, \forall g \in {\sf G}$, the fundamental K\"ahler 
potentials (\ref{bg23a}) are transformed as
\begin{equation}\label{bg23c}
K^j(gz, \overline{gz})= K^j(z, \bar z)+ \Phi^j(z, g) +
\overline{\Phi^j(z, g)},
\end{equation}
where $\Phi^j(z, g)$ are locally holomorphic functions of $z^{\alpha},
 \;\;\alpha=1, \dots, \frac{n-r}{2}$. These functions must satisfy
the following cocycle condition,
\begin{equation}
\Phi^j(z,g_2g_1)=\Phi^j(g_2z,g_1)+\Phi^j(z,g_2),\;\;\forall g_1,g_2\in{\sf G}
\end{equation}
which results from the group property $z \rightarrow g_2(g_1z)=(g_2g_1)z$.
\section*{IV. Expression of Berry connection in terms of the 
fundamental K\"ahler  potentials.}
With all the preliminary steps completed, we are in a position to
formulate the main result of this work.
\begin{quote}
{\bf Theorem}. Suppose that the cyclic adiabatic evolution of the 
dominant weight vector eigenket $\psi_{\bf l}$ defined by Eq.(\ref{bg2a})
is determined by the Shr\"odinger equation (Eq.\ref{bg3}). Let the 
Hamiltonian parameter space be a compact homogeneous K\"ahler manifold
${\sf G/H}$, where ${\sf G}$ is the compact evolution group, and 
${\sf H}$ its Cartan subgroup. Then the geometrical phase factor $\Omega$
acquired by the quantum state $\psi_{\bf l}$ is 
\begin{equation}\label{ex1}
\Omega=\int\limits^1_0 A_s ds,
\end{equation}
where the Berry connection $A_s$ is completely determined in terms of 
the fundamental K\"ahler potentials of the parameter space ${\sf G/H}$.
Explicitly, when the local complex parametrization 
$\left\{z^{\alpha}, \overline{z^{\alpha}}, \;\; \alpha=1, \dots, 
\frac{n-r}{2} \right\}$
on ${\sf G/H}$ is introduced, 
\begin{equation}\label{ex2}
A_s={\bf l}\cdot{\bf A}(z, \bar z), \;\;\;
{\bf A}(z, \bar z)={\cal L}_{z, \bar z}
\mbox{\boldmath $\kappa$}(z, \bar z).
\end{equation}
${\bf A}(z, \bar z)$ and $\mbox{\boldmath $\kappa$}(z, \bar z)$ are the
vectors in the root space of the Lie algebra of ${\sf G}$ given in the
orthonormal basis $\{ {\bf x}_j, \;\; j=1, \dots, r \}$
\begin{equation}\label{ex3}
\mbox{\boldmath $\kappa$}(z, \bar z)= \sum\limits^r_{j=1}\kappa^j(z, \bar z)
{\bf x}_j, \;\;\;
{\bf A}(z, \bar z)= \sum\limits^r_{j=1}A^j(z, \bar z){\bf x}_j.
\end{equation} 
${\cal L}_{z, \bar z}$ is the (hermitian) differential operator:
\begin{equation}\label{ex4}
{\cal L}_{z, \bar z}= i\sum\limits^{\frac{n-r}{2}}_{\alpha, \bar\alpha =1}
\left(\dot z^{\alpha}\partial_{\alpha} - \dot z^{\bar\alpha}
\partial_{\bar\alpha}\right),
\end{equation}
where $\dot z \equiv \frac{dz(s)}{ds}, \;\; 
z^{\bar\alpha}\equiv \overline{z^{\alpha}}, \;\;
\partial_{\alpha} \equiv \frac{\partial}{dz^{\alpha}},\;\;
\partial_{\bar\alpha} \equiv \frac{\partial}{dz^{\bar\alpha}}.$
The real functions $\kappa^j(z, \bar z)$ define the Cartan subgroup
element $k(z, \bar z) \in {\sf H}$ (Eq.\ref{bg18}) under (left) 
Mackey decomposition of a representative of the coset space ${\sf G/H}$
(Eqs.\ref{bg15}-\ref{bg18}). These functions are lineary connected with 
the fundamental K\"ahler potentials $K^j(z, \bar z)$ (Eq.\ref{bg21}).
\end{quote}

{\bf Proof:} The geometric phase factor $\Omega$ is given by the scalar 
product of the dominant weight vector ${\bf l}$ and the vector 
$\mbox{\boldmath ${\cal Q}$}$ (Eq.\ref{bg14}). The coordinates of the 
vector $\mbox{\boldmath ${\cal Q}$}$ in the root space of the Lie algebra
${\cal G}$ corresponding to the evolution group ${\sf G}$ are 
determined by Eq.\ref{bg11b}. When the local complex parametrization on 
the coset space ${\sf G/H}$ is introduced, the components ${\cal Q}^j$
of the vector $\mbox{\boldmath ${\cal Q}$}$ may be represented as a sum
of two integrals:
\begin{equation}\label{ex5}
{\cal Q}^j = -i\sum^{\frac{n-r}{2}}_{\alpha, \bar\alpha =1}\left\{
\int\limits_C{\rm Tr}\left( 
g^{-1}_1\partial_{\alpha}g_1 
h_j \right) dz^{\alpha} + \int\limits_C{\rm Tr}\left(g^{-1}_i
\partial_{\bar \alpha}g_1h_j \right) dz^{\bar\alpha}\right\}.
\end{equation}
The group element $g_1(z, \bar z)$ which is the representative of the
coset space ${\sf G/H}$ is decomposed (see Section 3) as 
\begin{equation}\label{ex6}
g_1(z, \bar z)= u(z)v^+(z, \bar z)k(z, \bar z), 
\end{equation}
where $u(z)\in {\sf B}_+, \;\; v^+(z, \bar z) \in {\bf B}_-,\;\; 
k(z, \bar z) \in {\sf H}$ are given by Eqs.(\ref{bg16}), (\ref{bg18}).
Let us recall that
\begin{displaymath}
k(z, \bar z)=\exp \left(\sum\limits_{j=1}^{r} \kappa^j(z, \bar z)h_j\right).
\end{displaymath}
It may be shown (see, for example, Itoh, Kugo and Kunimoto\cite{itoh})
that 
\begin{equation}\label{ex7}
{\rm Tr}\left(g^{-1}_1\partial_{\alpha}g_1 h_j \right)=
-\partial_{\alpha}\kappa^j(z, \bar z), \;\;\;
{\rm Tr}\left( g^{-1}_1 \partial_{\bar\alpha}g_1 h_j \right)=
\partial_{\bar\alpha}\kappa^j(z, \bar z).
\end{equation}
Indeed, noting that 
\begin{equation}\label{ex8}
\partial_{\bar\alpha}g_1(z, \bar z)= u(z)\partial_{\bar\alpha}
\Big(v^+(z, \bar z)k(z, \bar z)\Big),
\end{equation}
we find that
\begin{equation}\label{ex9}
{\rm Tr}\left( g^{-1}_1\partial_{\alpha}g_1 h_j \right) =
{\rm Tr}\Big( h_j(v^+(z, \bar z))^{-1}\partial_{\bar\alpha}
(v^+(z, \bar z))\Big) + 
\sum\limits^r_{i=1}\partial_{\alpha}(\kappa^i(z, \bar z))
{\rm Tr}(h_i h_j).
\end{equation}
The expression $\Big(v^+(z, \bar z)\Big)^{-1} \partial_{\bar\alpha}
\Big(v^+(z, \bar z)\Big)$ produces only terms belonging to the Borel 
subalgebra ${\cal B}_-$. As a consequence, the first term in 
the above equation equals to zero. Using the orthogonality condition 
for the Cartan 
subalgebra canonic basis elements 
\begin{equation}\label{ex9a}
{\rm Tr}(h_i h_j)=\delta_{ij},
\end{equation}
 we obtain
\begin{equation}\label{ex10}
{\rm Tr}\left( g^{-1}_1 \partial_{\bar\alpha}g_1 h_j \right)=
\partial_{\bar\alpha}\kappa^j(z, \bar z).
\end{equation}
In order to prove the first equation in (\ref{ex7}) we use
\begin{eqnarray}\label{ex11}
g^{-1}_1\partial_{\alpha}g_1&=& 
g^+_1\partial_{\alpha}\left(g^+_1\right)^{-1}
= k^+ v u^+\partial_{\alpha}\Big( (u^+)^{-1}v^{-1}(k^+)^{-1}\Big)
\nonumber\\
&=& k^+ v\partial_{\alpha}\Big( v^{-1}(k^+)^{-1}\Big),
\end{eqnarray}
Afterwards, we proceed with the proof as in the previous case. 

From Eqs.(\ref{ex5}), (\ref{ex7}) we find 
\begin{equation}\label{ex12}
{\cal Q}^j=i\sum\limits^{\frac{n-r}{2}}_{\alpha, \bar\alpha =1}\int\limits_0^1
\left(\dot z^{\alpha}\partial_{\alpha}-
\dot z^{\bar \alpha}\partial_{\bar\alpha}\right) \kappa^j(z, \bar z)ds
\equiv \int\limits^1_0 A^j(z, \bar z) ds.
\end{equation}
Thus, the vector in the root space of ${\cal G}$ 
\begin{equation}\label{ex13}
{\bf A}(z, \bar z)=i\sum\limits^{\frac{n-r}{2}}_{\alpha, \bar\alpha =1}
\left(\dot z^{\alpha}\partial_{\alpha}-
\dot z^{\bar \alpha}\partial_{\bar\alpha}\right) \mbox{\boldmath $\kappa$}
(z, \bar z) \equiv {\cal L}_{z, \bar z}\mbox{\boldmath $\kappa$}
(z, \bar z)
\end{equation}
is introduced, the Berry connection and the Berry geometrical phase
factor are determined by Eqs.(\ref{ex1}), (\ref{ex2}) respectively. 
$\Box$

In the next section we use this result to demonstrate that 
the holomorphic action of 
the evolution group ${\sf G}$ on the Hamiltonian parameter space
${\sf G/H}$  induces the gauge transformation of the Berry potentials. 
In addition, the Berry curvature and the Berry geometrical phase factor will be
obtained.
\section*{V. Gauge transformation and Berry curvature}
Consider the transformation of the vector ${\bf A}(z, \bar z)$ under the
holomorphic action of the group ${\sf G}$ on the homogeneous K\"ahler
${\sf G/H}$. As soon as the fundamental K\"ahler potentials 
$K^j(z, \bar z), \;j=1, \dots, r$ are transformed in accordence with 
Eq.(\ref{bg23c}), the vector $\mbox{\boldmath $\kappa$}(z, \bar z)$
changes in a similar fashion, i.e.
\begin{equation}\label{g1}
\mbox{\boldmath $\kappa$}(z, \bar z) \rightarrow 
\mbox{\boldmath $\kappa$}(gz, \overline{gz})=\mbox{\boldmath $\kappa$}
(z, \bar z) + \mbox{\boldmath $\phi$}(g,z)+
\overline{\mbox{\boldmath $\phi$}(g, z)}.
\end{equation}
Indeed, given the decomposition of the coset space representative 
(Eqs.\ref{bg15}-\ref{bg18}), the action of an arbitrary group element 
$g_2\in {\sf G}$ on the coset space ${\sf G/H}$ is defined 
(by Coleman, Wess and Zumino\cite{coleman}) as
\begin{equation}\label{g2}
g_2 u(z)=u(g_2z)g_-(z, g_2),
\end{equation}
and $g_2z$ is a rational function of $z$. 
Once the nonlinear realization of the group
action on the coset space ${\sf G/H}$ is determined (Eq.\ref{g2}),
the the transformation law (\ref{g1}) may be proved using 
the Mackey-type decomposition of the product $g_1\cdot g_2$ (where
$g_1 \in {\sf G/H}$ is given by 
Eqs.(\ref{bg15})-(\ref{bg18}))(for further details see 
Itoh, Kogo and Kunimoto\cite{itoh}.
The change of the real vector $\mbox{\boldmath $\kappa$}(z, \bar z)$
(Eq.\ref{g1}) under the holomorphic action of the group on its coset
space leads to the gauge transformation of the vector ${\bf A}(z, \bar z)$:
\begin{equation}\label{g3}
{\bf A}(z, \bar z) = {\cal L}_{z, \bar z}
\mbox{\boldmath $\kappa$}(z, \bar z) \rightarrow
{\bf A}(gz, \overline{gz})= {\bf A}(z, \bar z)+
d {\bf W}(z, \bar z),
\end{equation}
where the real vector ${\bf W}(z, \bar z)$ is defined in terms of the
vectors $\mbox{\boldmath $\phi$}(g, z), \; 
\overline{\mbox{\boldmath $\phi$}(g, z)}$:
\begin{equation}\label{g4}
{\bf W}(z, \bar z)\equiv i\Big( \mbox{\boldmath $\phi$}(g, z)-
\overline{\mbox{\boldmath $\phi$}(g, z)} \Big).
\end{equation}
Respectively, the abelian Berry connection $A_s$ defined by Eq.(\ref{ex2})
is transformed as
\begin{eqnarray}\label{g5}
A_s(z, \bar z)\rightarrow A_s(gz, \overline{gz})&=&A_s(z, \bar z)+
dW(z, \bar z) \nonumber\\
W(z, \bar z) & \equiv & {\bf l} \cdot {\bf W}(z, \bar z).
\end{eqnarray}
Note that the expression (\ref{ex1}) for $\Omega$ may be rewritten as
\begin{equation}\label{g7}
\Omega=\sum\limits^{\frac{n-r}{2}}_{\alpha=1} \int\limits_C A_{\alpha}
(z, \bar z)dz^{\alpha} + \sum\limits^{\frac{n-r}{2}}_{\alpha=1} 
\int\limits_C A_{\bar\alpha}(z, \bar z)dz^{\bar\alpha} 
\end{equation}
where 
\begin{eqnarray}\label{g8}
A_{\alpha}(z, \bar z)& \equiv & i \partial_{\alpha}
\Big({\bf l}\cdot \mbox{\boldmath $\kappa$} (z, \bar z)\Big),
\nonumber\\
A_{\bar\alpha}(z, \bar z)& \equiv & 
-i \partial_{\bar\alpha}
\Big({\bf l}\cdot \mbox{\boldmath $\kappa$} (z, \bar z)\Big).
\end{eqnarray}
Under the holomorphic action of the group ${\sf G}$ on the homogeneous
K\"ahler manifold ${\sf G/H}$, $A_{\alpha}(z, \bar z), \;
A_{\bar\alpha}(z, \bar z)$ transform as
\begin{eqnarray}\label{g9}
A_{\alpha}(z, \bar z)&\rightarrow & A_{\alpha}(gz, \overline{gz})=
A_{\alpha}(z, \bar z)+ i\partial_{\alpha} \Big({\bf l} 
\cdot \mbox{\boldmath $\phi$}(g,z)\Big)
\nonumber\\
A_{\bar\alpha}(z, \bar z)&\rightarrow & 
A_{\bar\alpha}(gz, \overline{gz})=
A_{\bar\alpha}(z, \bar z)- i\partial_{\bar\alpha} \Big({\bf l} 
\cdot \mbox{\boldmath $\bar\phi$}(g,z)\Big).
\end{eqnarray}
Using the Stokes theorem, we obtain the expression for the Berry geometrical
factor in terms of the surface integral,
\begin{equation}\label{g6}
\Omega=\int\limits_S F, \;\;\; 
F=\sum\limits^{\frac{n-r}{2}}_{\alpha, \bar\beta=1} 
\frac{\partial^2 K^{({\bf l})}(z, \bar z)}
{\partial z^{\alpha}\partial z^{\bar\beta}} dz^{\alpha} \wedge
dz^{\bar\beta},
\end{equation}
where
\begin{equation}\label{g10}
K^{({\bf l})}(z, \bar z)=2\Big({\bf l}\cdot 
\mbox{\boldmath $\kappa$}(z, \bar z)\Big),
\end{equation}
and $S$ is any oriented surface in the parameter space ${\sf G/H}$ with
$\partial S =C$. As it may be seen from Eqs.(\ref{g1}), (\ref{g6}),
 the Berry curvature $F$ is invariant under the gauge 
transformation (Eq.\ref{g5}) induced by the holomorphic group action on
 the parameter space ${\sf G/H}$.

A simple way to calculate the vector 
$\mbox{\boldmath $\kappa$}(z, \bar z)$ which determines the Berry connection
and the Berry curvature is to use Eq.(\ref{bg21}). This formula
connects the vector $\mbox{\boldmath $\kappa$}(z, \bar z)$ with the
fundamental K\"ahler potentials given by Eqs.(\ref{bg22})-(\ref{bg23b}).
\section*{VI. Examples of adiabatic evolutions induced by Lie groups}
\subsection*{$SU(2)$ adiabatic evolution}
The space $S^2= SU(2)/U(1)$ is the simplest homogeneous K\"ahler manifold. The
 generators of the group $SU(2)$ are the Pauli matrices:
\begin{eqnarray}\label{bg39}
\sigma_1 = \left(
\begin{array}{cc}
0 & 1 \\
1 & 0
\end{array}
\right); \;\;
\sigma_2 = \left(
\begin{array}{cc}
0 & -i \\
i & 0 
\end{array} 
\right); \;\;
\sigma_3 = \left(
\begin{array}{cc}
1 & 0\\
0 & -1
\end{array}
\right).
\end{eqnarray}
The elements of the canonical basis of the Lie algebra $SU(2)$ in the
fundamental (spinor) representation are
\begin{equation}\label{bg40}
E_1=\frac{1}{2}(\sigma_1 +i\sigma_2); \;\;
E_2=\frac{1}{2}(\sigma_1 -i\sigma_2); \;\;
H=\frac{1}{\sqrt{2}}\sigma_3.
\end{equation}
Correspondingly, the canonical commutation relations for the $SU(2)$ Lie 
algebra have the following form:
\begin{equation}\label{bg41}
[e_1, e_2]= \sqrt{2} h, \;\; [h, e_1] =\sqrt{2} e_1, \;\; 
[h, e_2]=-\sqrt{2}e_2. 
\end{equation}
The rank of the Lie algebra $SU(2)$
is equal to one, $r=1$. The root space is one-dimensional, $R^1$, with
two opposite roots $\pm \mbox{\boldmath $\alpha$}= \pm\sqrt{2} {\bf x}$,
and the primitive root is 
$\mbox{\boldmath $\gamma$}= \mbox{\boldmath $\alpha$}$.
The fundamental weight vector  $\mbox{\boldmath $\omega$}$ is determined by
Eq.(\ref{bg1a}), 
$\mbox{\boldmath $\omega$}= \frac{1}{2}\mbox{\boldmath $\alpha$}$.
An arbitrary irreducible representation is  defined by the dominant
weight vector,
\begin{equation}\label{bg42}
{\bf l}= l \mbox{\boldmath $\omega$}=\frac{l}{\sqrt{2}}{\bf x}, \;\; 
l=0, \pm 1, \pm 2, \dots . 
\end{equation}
We consider the cyclic adiabatic evolution of the dominant weight vector
 eigenket $\psi_{{\bf l}}$ which is the eigenvector of the following 
eigenvalue problem: 
\begin{equation}\label{bg43}
h\psi_{{\bf l}}=\frac{l}{\sqrt{2}} \psi_{{\bf l}}, \;\; l = 0, \pm 1, \pm 2, 
\dots . 
\end{equation}
In accordance with Eqs.(\ref{bg14}), (\ref{bg11}) after the cyclic 
adiabatic evolution the state $\psi_{{\bf l}}$
acquires the geometrical phase factor $\Omega$ 
(Eqs.\ref{bg11b}, \ref{bg14a}).
The abelian Berry connection $A_s$ is given by (\ref{ex2}). The vector
$\mbox{\boldmath $\kappa$}$ in the case of $SU(2)$ evolution group 
may be obtained by the (left) Mackey-type decomposition of 
a representative $g_1(z,\bar z)$ of the coset space $SU(2)/U(1)$:
\begin{equation}\label{bg45}
g_1(z, \bar z) = u(z)g_-(z, \bar z), \;\;
u(z) = \exp(z e_1).
\end{equation}
The element $g_-(z, \bar z)$ is found in accordance with the general procedure
 described in Section III:
\begin{eqnarray}\label{bg46}
g_-(z, \bar z)&=&\exp \Big((y(z, \bar z) e_2 \Big) \exp\Big(\kappa(z, \bar z)
h \Big) \nonumber\\
y(z, \bar z)&=& -\frac{\bar z}{1+z\bar z}, \;\;
\kappa(z, \bar z)= \frac{1}{\sqrt{2}}\ln (1+ z \bar z).
\end{eqnarray}
Therefore the vector (in the root space of the Lie algebra $su(2)$)
$\mbox{\boldmath $\kappa$}$  is given by
\begin{equation}\label{bg46a}
\mbox{\boldmath $\kappa$}(z, \bar z)= \frac{1}{\sqrt{2}}\ln(1+ z\bar z)
{\bf x},
\end{equation}
and the abelian Berry connection $A_s$ is expressed in terms of the complex
coordinates of the coset space $SU(2)/U(1)$ as:
\begin{equation}\label{bg46b}
A_s= \frac{il}{2}\left( \frac{\dot z \bar z - \dot{\bar z} z}{1+ z \bar z}
\right).
\end{equation}
Using the Stokes theorem we determine
the Berry  geometrical phase factor  
(Eqs.\ref{bg11b}, \ref{bg14a}):
\begin{equation}\label{bg49}
\Omega = -i l\int\!\!\!\int \frac{dz \wedge d\bar z}{(1+ z\bar z)^2}, 
\;\;\; l= 0, \pm 1, \pm 2, \dots .
\end{equation}
In  order to compare Eq.(\ref{bg49}) with the original Berry result\cite{berry}
for the spin procession  in the time-dependent magnetic field, use  the 
stereographic  projection of the two-dimensional sphere  with the 
unit radius : 
\begin{equation}\label{bg50}
|z| = \cot \theta/2  \quad \arg z =\varphi.
\end{equation}
Then the geometrical phase factor is equal to
\begin{equation}\label{bg51}
\omega = \frac{l}{2}\int\!\!\!\int \sin\theta d\theta \wedge  d\varphi,
\quad
l=0, \pm 1, \pm 2, \dots .
\end{equation}
It is this result that was obtained by Berry\cite{berry}.
\subsection*{$SU(3)$ adiabatic evolution}
The Cartan subgroup of the $SU(3)$ group is $U(1) \times U(1)$, so the Berry
geometrical   phase factor will be determined by the geometry of the Flag
 manifold $SU(3)/U(1)\times U(1)$. The canonical basis  of $SU(3)$ Lie algebra is
introduced with the help of the eight Gell-Mann generators:
\begin{equation}\label{bg52}
\begin{array}{ccc}
\lambda_1 = \left(
\begin{array}{ccc}
0 & 1 & 0\\
1 & 0 & 0\\
0 & 0 & 0
\end{array}
\right) &
\lambda_2 = \left(
\begin{array}{ccc}
0 & -i & 0\\
i & 0 & 0\\
0 & 0 & 0
\end{array}
\right) &
\lambda_3 = \left(
\begin{array}{ccc}
1 & 0 & 0\\
0 & -1 & 0\\
0 & 0 & 0
\end{array}
\right) 
\nonumber\\
\lambda_4 = \left(
\begin{array}{ccc}
0 & 0 & 1\\
0 & 0 & 0\\
1 & 0 & 0
\end{array}
\right) &
\lambda_5 = \left(
\begin{array}{ccc}
0 & 0 & -i\\
0 & 0 & 0\\
i & 0 & 0
\end{array}
\right) &
\lambda_6 = \left(
\begin{array}{ccc}
0 & 0 & 0\\
0 & 0 & 1\\
0 & 1 & 0
\end{array}
\right)
\nonumber\\
\lambda_7 = \left(
\begin{array}{ccc}
0 & 0 & 0\\
0 & 0 & -i\\
0 & i & 0
\end{array}
\right) &
\lambda_8 = \frac{1}{\sqrt{3}} \left(
\begin{array}{ccc}
1 & 0 & 0\\
0 & 1 & 0\\
0 & 0 & -2
\end{array}
\right).&
\end{array}
\end{equation} 
Then the elements of the canonical Cartan-Weyl basis in the 
fundamental three-dimensional representation are given by
\begin{equation}\label{bg53}
\begin{array}{ccccccccc}
E_{12} &=& 1/2 (\lambda_1 + i\lambda_2); & 
E_{23} &=& 1/2 (\lambda_6 + i\lambda_7); & 
E_{13} &=& 1/2 (\lambda_4 + i\lambda_5);  
\\
E_{21} &=& 1/2 (\lambda_1 - i\lambda_2); & 
E_{32} &= &1/2 (\lambda_6 - i\lambda_7); & 
E_{31} &=& 1/2 (\lambda_4 - i\lambda_5); 
\\
H_1 &=& \frac{\sqrt{3}}{2}\lambda_3 + \frac{\lambda_8}{2};&
H_2 &=& -\frac{\lambda_3}{2}+\frac{\sqrt{3}}{2}\lambda_8. & 
\end{array}
\end{equation}
The rank of the Lie algebra $SU(3)$ is equal to two, $r=2$. The root space is
two dimensional,  $R^2$, and the canonical commutation relations
 determining the
positive root vectors are
\begin{equation}\label{bg54}
\begin{array}{ccccccccc}
\lbrack h_1, e_{12}\rbrack &=& \frac{3}{\sqrt{6}} e_{12}&; &
\lbrack h_2, e_{12} \rbrack &=& -\frac{1}{\sqrt{2}}e_{12} &;
\\
\lbrack h_1, e_{13}\rbrack &=& \frac{3}{\sqrt{6}}e_{13} &;&
\lbrack h_2, e_{13} \rbrack &=& \frac{1}{\sqrt{2}} e_{13}&;
\\
\lbrack h_1, e_{23}\rbrack &=& 0&;& 
\lbrack h_2, e_{23} \rbrack &=& -\sqrt{2}e_{23}&.
\end{array}
\end{equation}
From the commutation relations (\ref{bg54}) we find six non-zero
root vectors, 
\begin{equation}\label{bg55}
\pm \mbox{\boldmath $\alpha_1$}=
\pm \left( \frac{3}{\sqrt{6}}; -\frac{1}{\sqrt{2}}\right), \;\;\;
\pm \mbox{\boldmath $\alpha_2$}=
\pm \left( \frac{3}{\sqrt{6}};\frac{1}{\sqrt{2}}\right),\;\;\;
\pm \mbox{\boldmath $\alpha_3$}= \left(0; -\sqrt{2}\right).
\end{equation}
The root diagram is a hexagon, the two primitive roots are 
$\mbox{\boldmath $\gamma_1$}=\mbox{\boldmath $\alpha_3$}$,
$\mbox{\boldmath $\gamma_2$}=\mbox{\boldmath $\alpha_2$}$.
The fundamental weight are found to be
\begin{equation}\label{bg56}
\mbox{\boldmath $\omega_1$}= \left(\frac{1}{\sqrt{6}};
-\frac{1}{\sqrt{2}} \right) ,\;\;\;
\mbox{\boldmath $\omega_2$}= \left(\frac{2}{\sqrt{6}}; 0 \right).
\end{equation}
An arbitrary irreducible representation is defined by the (two-dimensional)
dominant weight vector:
\begin{equation}\label{bg57}
{\bf l}= l_1\mbox{\boldmath $\omega_1$}+ l_2 
\mbox{\boldmath $\omega_2$}, \;\;\;
l_1, l_2 = 0, \pm 1, \pm 2, \dots .
\end{equation}
The coordinates of this vector in the root space are:
\begin{equation}\label{bg58} 
\tilde l_1= \frac{l_1}{\sqrt{6}}+ \frac{2 l_2}{\sqrt{6}}, \;\;\;
\tilde l_2=-\frac{l_1}{\sqrt{2}}.
\end{equation}
The dominant weight vector eigenket $\psi_{{\bf l}}$ is an eigenvector 
of both Cartan Lie algebra elements $h_1, h_2$ with eigenvalues 
$\tilde l_1, \tilde l_2$. 
It follows from the Theorem (Section 4) that in order to find the 
geometric phase factor $\Omega$ acquired by
the dominant weight vector eigenket $\psi_{\bf l}$, one should determine
the two-component vector ${\bf k}(z, \bar z)$ in the root space of the
Lie algebra $su(3)$. As it may be seen from Eq.(\ref{bg21})
the components of this vector in the orthogonal basis
of the root space are linear combinations of the fundamental K\"ahler
potentials $K^1(z, \bar z),\; K^2(z, \bar z)$ of the homogeneous 
K\"ahler manifold $SU(3)/U(1)\times U(1)$. The fundamental K\"ahler
potentials $K^1(z, \bar z),\; K^2(z, \bar z)$ are constructed using
Eq.(\ref{bg23b}), where the element $u(z)$ is taken in the 
fundamental $3 \times 3$ representation:
\begin{equation}\label{bg58a}
u(z)\equiv \exp \Big(z_1 E_{12}+ z_2 E_{23}+ z_3 E_{13} \Big)
= \left(
\begin{array}{ccc}
1 & z_1 & z^+_3\\
0 & 1 & z_2\\
0 & 0 & 1
\end{array}
\right). 
\end{equation}
We have denoted $z^{\pm}= z_3 \pm \frac{1}{2}z_1z_2$, and the projection
matrices found from Eqs.(\ref{bg22}) are
\begin{equation}\label{bg61}
\eta_1= \left(
\begin{array}{ccc}
0 & 0 & 0\\
0 & 1 & 0\\
0 & 0 & 1
\end{array}
\right) \;\;\;
\eta_2= \left(
\begin{array}{ccc}
0 & 0 & 0\\
0 & 0 & 0\\
0 & 0 & 1
\end{array}
\right),
\end{equation}
Knowing $u(z), \eta_1, \eta_2$ in the fundamental representation,
the fundamental K\"ahler potentials are calculated, and we obtain
\begin{equation}\label{bg60}
K^1(z, \bar z)= \ln(1+z_1 \bar z_1 + z_3^- \bar z_3^-), \;\;\;
K^2(z, \bar z)=\ln(1+z_2 \bar z_2 + z_3^+ \bar z_3^+).
\end{equation}
Finding the coefficients ${\rm Tr}(h_j \eta_i)$ (in the fundamental 
represenation $h_1 \equiv H_1, \; h_2 \equiv H_2$), we get the components
of the vector $\mbox{\boldmath $\kappa$} (z, \bar z)$ in the root space
of $su(3)$ algebra:
\begin{equation}\label{bg62}
\kappa^1(z, \bar z)=\frac{\sqrt{6}}{4}K^1(z, \bar z), \;\;\;
\kappa^2(z, \bar z)=\frac{1}{\sqrt{2}}K^2(z, \bar z)- \frac{1}{2\sqrt{2}}
K^1(z, \bar z). 
\end{equation} 
(Note that the expressions for the components ofthe vector
$\mbox{\boldmath $\kappa$} (z, \bar z)$ in terms of the complex coordinates
Eqs.(\ref{bg60}), (\ref{bg62}) may be obtained also by the (left) Mackey
decomposition of the coset space representative $g_1$,
 Eqs.(\ref{bg15})-(\ref{bg18})).

For the case of $SU(3)$ evolution group the Berry connection $A_s$ 
given by Eq.(\ref{ex2}) is (\ref{bg62a}):
\begin{eqnarray}\label{bg62a}
A_s &=& {\cal L}_{z, \bar z} \left\{{\bf l}\cdot 
\mbox{\boldmath $\kappa$} (z, \bar z)\right\}, \;\;
{\bf l}\cdot \mbox{\boldmath $\kappa$}(z, \bar z) =\frac{1}{2}
\Big[(l_1+l_2)K^1(z, \bar z)-l_1 K^2(z, \bar z)\Big]
\nonumber\\
{\cal L}& =& \sum\limits^3_{\alpha, \bar\alpha=1} \left(
\dot z^{\alpha}\frac{\partial}{\partial z^{\alpha}} -
\dot z^{\bar\alpha} \frac{\partial}{\partial z^{\bar\alpha}} \right),
\end{eqnarray}
and the Berry curvature $F$ (Eq.\ref{g6}) is
\begin{equation}\label{bg62b}
F=\sum\limits^3_{\alpha,\bar\beta=1}
\frac{\partial^2 K^{({\bf l})}(z, \bar z)}
{\partial z^{\alpha}\partial z^{\bar\beta}} dz^{\alpha}\wedge
dz^{\bar\beta},
\end{equation}
where the real function $K^{({\bf l})}(z, \bar z)$ is a linear 
combination of the fundamental K\"ahler potentials 
$K_1(z, \bar z), \;\; K_2(z, \bar z)$ with the integer coefficients: 
\begin{equation}\label{bg64}
K^{({\bf l})}(z, \bar z)= (l_1 + l_2)K^1(z, \bar z) - l_1 K^2(z, \bar z), \;\;\;
l_1, l_2 = 0, \pm1, \pm2 \dots .
\end{equation}
Note that a common approach to the $SU(3)$ group evolution is 
to use the
Euler coordinates that are similar to the Euler angle parameters of
$SU(2)$. Such method has been
used by Byrd\cite{byrd}, Arvind, Mallesh and Mukunda\cite{arvind},
Khanna, Mukhopadhaya, Simon, and Mukunda\cite{khanna} in connection with the
evolution of a three level system.
While the geometric phase factor for this case has been found, none of
them noticed that in cases under their consideration
there exists an intimate relation between the
geometric phase factors, Berry connections and Berry curvature and the
fundamental K\"ahler potentials of parameter spaces. 
\section*{VII. Conclusions.}
In this paper we have considered the adiabatic evolution determined
by a compact Lie group evolution operator taken in an 
arbitrary irreducible representation. 
It has been demonstrated that when the parameter space of
the Hamiltonian is a homogeneous K\"ahler manifold its fundamental 
K\"ahler potentials completely determine Berry geometrical phase 
factor. Besides, we have shown that Berry geometrical factor 
and the Berry connections depend on a set of integers a number of 
which equals to the rank of the corresponding 
Lie algebra. These integers determine irreducible 
representation in which quantum states form a basis.
\subsection*{Acknowledgements.}
For valuable comments and discussions that have contributed
to  this work , many thanks to  J. Avron, and A. Nepomnyashiy. 

\end{document}